\documentclass[journal=apchd5]{achemso}
\usepackage[version=3]{mhchem} 
\usepackage[T1]{fontenc}       

\usepackage{amsmath,amssymb}
\usepackage{epsfig}
\usepackage{graphicx}
\usepackage{dcolumn}
\usepackage{bm}
\usepackage{mathtools}
\usepackage{hyperref}
\usepackage{color}
\usepackage[a4paper]{anysize}

\newcommand{\e}{{\bm e}}
\newcommand{\A}{{\bm A}}
\newcommand{\E}{{\bm E}}
\newcommand{\B}{{\bm B}}
\newcommand{\Er}{{\bm{\mathcal E}}}
\newcommand{\Br}{{\bm{\mathcal B}}}
\newcommand{\C}{{\mathcal C}}
\renewcommand{\P}{{\bf P}}
\renewcommand{\S}{{\bf S}}

\newcommand{\p}{{\bm p}}
\newcommand{\m}{{\bm m}}

\renewcommand{\j}{{\bm j}}
\renewcommand{\k}{{\bm{k}}}
\newcommand{\q}{{\bm{q}}}
\newcommand{\br}{{\bm{r}}}

\def\gsim{\lower.35em\hbox{$\stackrel{\textstyle>}{\textstyle\sim}$}}
\def\lsim{\lower.35em\hbox{$\stackrel{\textstyle<}{\textstyle\sim}$}}

\title{Plasmon-enhanced near-field chirality in twisted van der Waals heterostructures} 
\author{T. Stauber} 
\email{tobias.stauber@csic.es}
\affiliation{Departamento de Teor\'{\i}a y Simulaci\'on de Materiales, Instituto de Ciencia de Materiales de Madrid, CSIC, 28049 Madrid, Spain}
\affiliation{Institute for Theoretical Physics, University of Regensburg, D-93040 Regensburg, Germany}
\author{T. Low}
\affiliation{Department of Electrical \& Computer Engineering, University of Minnesota, Minneapolis, Minnesota 55455, USA}
\author{G. G\'omez-Santos}
\affiliation{Departamento de F\'{\i}sica de la Materia Condensada, INC and IFIMAC, Universidad Aut\'onoma de Madrid, E-28049 Madrid, Spain}
\begin{document}
\begin{abstract}
It is shown that chiral plasmons, characterized by a longitudinal magnetic moment accompanying the longitudinal charge plasmon, lead to electromagnetic near-fields that are also chiral. For twisted bilayer graphene, we estimate that the near field chirality of screened plasmons can be several orders of magnitude larger than that of the related circularly polarized light.
The chirality also manifests itself in a deflection angle that is formed between the direction of the plasmon propagation and its Poynting vector.
Twisted van der Waals heterostructures might thus provide a novel
platform to promote enantiomer-selective physio-chemical processes in chiral molecules without the application of a magnetic field or external nano-patterning that break time-reversal, mirror plane or inversion symmetry, respectively. 
\end{abstract}
\section*{Introduction} 
Chirality is an important aspect in life as only one enantiomer of amino acids is present in nature.\cite{Barron04} Furthermore, chiral objects can only be distinguished through the interaction with other chiral objects. A prominent example is the circular dichroism of chiral molecules where a different absorption cross section is seen when changing the chirality of the incident light.\cite{Kuwata05,Rogacheva06,Govorov10,Guerrero11,Zhou12} This  gives rise to an asymmetry factor denoting the normalised difference between the two absorption cross sections. 

Exposing (chiral) organic molecules to circularly polarized light (CPL) might lead to modified chemical reactions, but in general the asymmetry factor is of the order of $10^{-3}$ and thus negligible. This is related to the fact that the length scales between the two interacting chiral objects usually differ by several orders of magnitudes, i.e., for CPL $ak\approx10^{-3}$ in the optical regime where $k=2\pi/\lambda$ denotes the wave number and $a$ the length scale of the molecule. 

There have been proposals to enhance the enantioselectivity in the excitation of chiral molecules by superchiral light.\cite{Tang11} Also chiral metamaterials and plasmonics show promising results,\cite{Plum09,Guerrero11,Hentschel17} and even in Bernal-stacked bilayer graphene axial coupling can be induced.\cite{Kammermeier19} But with the advent of atomically thin two-dimensional (2D) crystals,\cite{Geim13} new opportunities arise and macroscopically large chiral objects with inherent uniaxial coupling can be designed following a bottom-up approach by placing the 2D crystals on top of each other with a rotational mismatch.  

The circular dichroism of twisted bilayer graphene (TBG) was first observed by Kim et al.,\cite{Kim16} but is rather weak. It could be increased by stacking multiple layers with a definite relative twist angle on top of each other. For $n$ layers, the intrinsic length scale given by the interlayer separation $a\sim3.4\mathring{\text{A}}$ would be increased by $a\rightarrow a^*=na$ and the dimensionless chirality scale $a^*k$ could reach the order of unity.\cite{Kim16} Another way to increase the dimensionless chirality scale $ak$ would be to decrease the wavelength of the chiral field. 

Reducing the wavelength of the electromagnetic field is possible using confined plasmonic modes. Adsorbed molecules naturally facilitates the coupling of far-field light into these plasmonic modes, and their excitation significantly enhances otherwise weak light-matter interactions\cite{low2014graphene}. 
 In fact, the plasmonic wave-length can be considerably decreased in graphene,\cite{Fei12,Chen12} which is related to the fact that the Fermi velocity $v_F$ is two orders of magnitude smaller the speed of light $c$.\cite{Wunsch06,Stauber14} We shall be guided by this approach and introduce a new degree of freedom in the form of the chirality of the near-field by considering plasmons in twisted van der Waals heterostructures.

\section*{Chiral plasmons in TBG}
Twisted bilayer graphene\cite{Lopes07,Shallcross08,Suarez10,Schmidt10,Li10,Trambly10,Bistritzer11,Dean13} (TBG) has attracted tremendous attention since the discovery of correlated insulator states\cite{Kim17,Cao18a} as well as superconductivity\cite{Cao18b,Yankowitz19,Lu19} close to the magic angle of $\theta\sim1.08^\circ$. Also, plasmonics in TBG has received considerable interest,\cite{Stauber13,Stauber14,Hu17,Lin20} and interband collective modes around the charge neutrality point have been predicted\cite{Stauber16} and observed.\cite{Hesp19} Furthermore, for systems with narrow band widths such as TBG, plasmons are expected to be long-lived\cite{Levitov19,Khaliji20} and for twist angles of less than half a degree, the possibility of a photonic-crystal for these collective excitations opens up.\cite{Sunku18} Scanned probe optical techniques can also be used to determine the local twist angle and domain structure.\cite{Sunku20}
\begin{figure}[t]
\includegraphics[width=0.99\columnwidth]{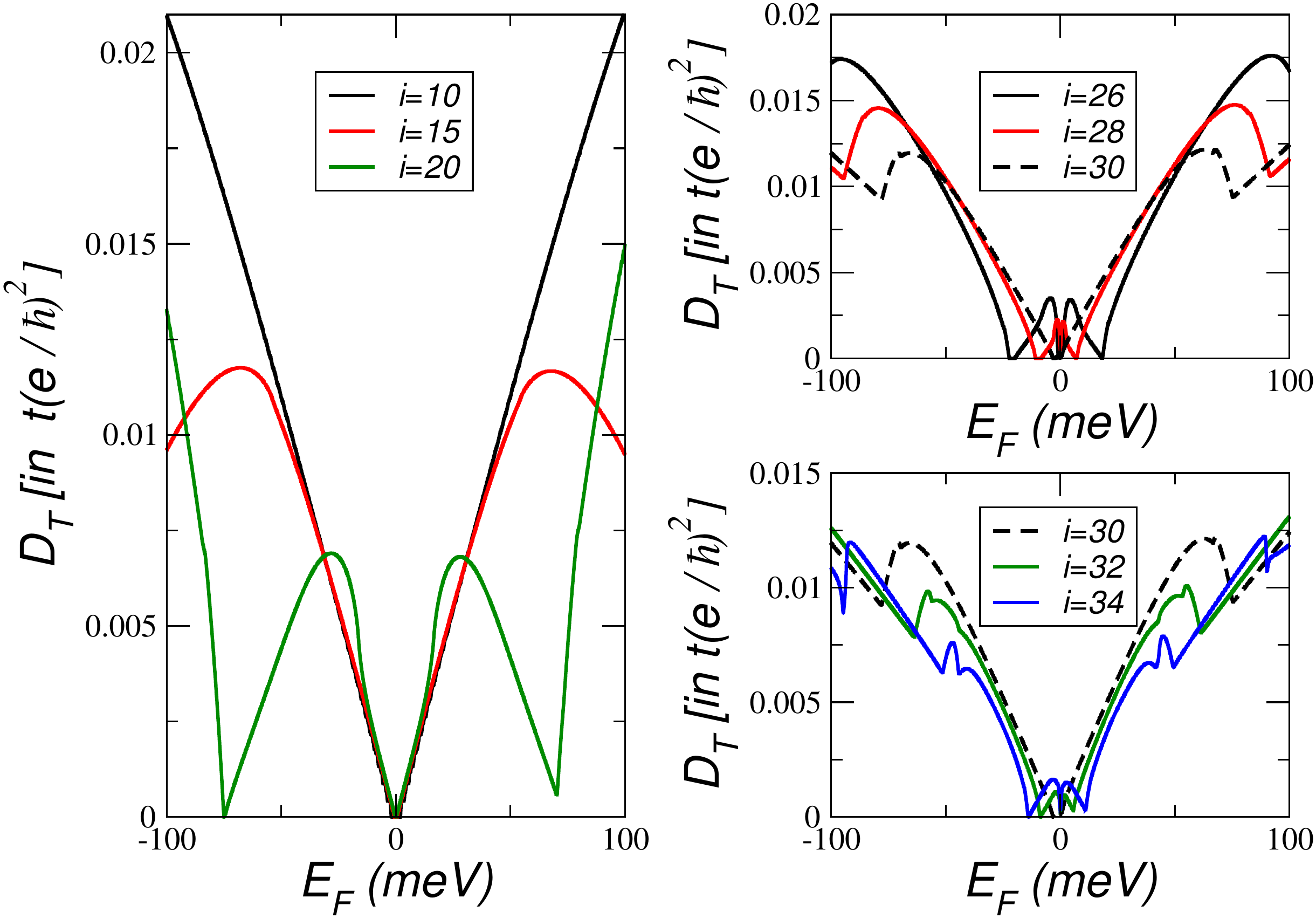}
\caption{\label{Fig1} The total Drude weight in units of $t(e/\hbar)^2$ for large twist angles $\theta_i=3.15^\circ,2.13^\circ,1.61^\circ$ with $i=10,15,20$ (left) and for small twist angles $\theta_i=1.25^\circ,1.16^\circ,1.08^\circ,1.02^\circ,0.96^\circ$ with $i=26,28,30,32,34$ (right) as function of the Fermi energy $E_F$. Commensurate twist angles are parametrised by $\cos\theta_i=\frac{3i^2+3i+1/2}{3i^2+3i+1}$ and $t=2.78$eV denotes the in-plane hopping parameter.}
\end{figure}
\begin{figure}[t]
\includegraphics[width=0.99\columnwidth]{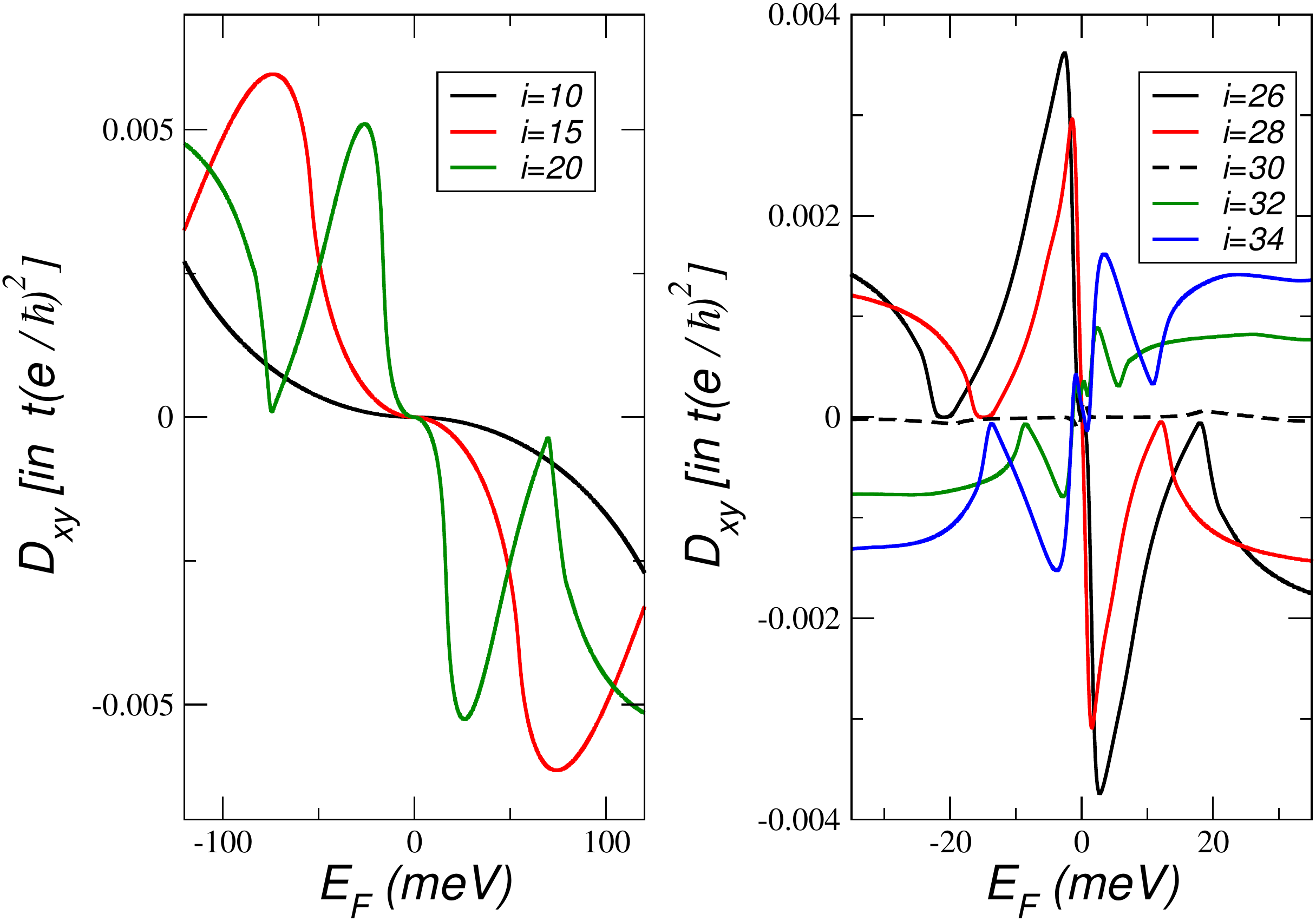}
\caption{\label{Fig2} The chiral Drude weight in units of $t(e/\hbar)^2$ for large twist angles $\theta_i=3.15^\circ,2.13^\circ,1.61^\circ$ with $i=10,15,20$ (left) and for small twist angles $\theta_i=1.25^\circ,1.16^\circ,1.08^\circ,1.02^\circ,0.96^\circ$ with $i=26,28,30,32,34$ (right) as function of the Fermi energy $E_F$. Commensurate twist angles are parametrised by $\cos\theta_i=\frac{3i^2+3i+1/2}{3i^2+3i+1}$ and $t=2.78$eV denotes the in-plane hopping parameter.}
\end{figure}

Like any metal layer, doped TBG hosts plasmons with dispersion given by $\omega^2=qD_T/(\epsilon_0(\epsilon_1+\epsilon_2))$. It depends on the dielectric constants of the upper ($\epsilon_1$) and lower ($\epsilon_2$) half-planes as well as on the total Drude weight or charge stiffness $D_T$ as defined in the Supplemental Information (SI). For 2D systems with dispersion relation $\epsilon_\k\sim|\k|^\nu$ and spin- and valley degeneracy $g_s$, $g_v$, it is proportional to the Fermi energy $E_F$ with $D_T=\frac{g_sg_v\nu E_F}{4\pi\hbar^2}$.\cite{Stauber14} This linear behavior is strongly modified in the case of TBG due to the appearance of mini-bands\cite{Stauber13} as shown in Fig. \ref{Fig1}. Placing a metal\cite{Principi11} or a large dielectric\cite{Stauber12} at a finite distance $d$ from the TBG sheet will effectively screen the plasmons, leading to the linear dispersion relation $\omega=v_sq$, characterized by the sound velocity $v_s$.

Chiral plasmonics relies on TBG's intrinsic chirality that gives rise to circular dichroism in the absence of symmetry-breaking fields.\cite{Kim16,Suarez17,Addison19,Ochoa20} It is due to the small, but finite separation of the two graphene layers which requires that even the minimal response theory has to be formulated within a $4\times4$ matrix as defined in the SI.\cite{Stauber18,Stauber18b,Bahamon20,Lin20} Chirality can then be discussed via the chiral Drude weight $D_{xy}$ that links the $x$-direction of layer 1 to the $y$-direction of layer 2. This quantity is shown in Fig. \ref{Fig2} and can be related to the vector-product of the sheet current densities as\cite{Stauber20}
\begin{align}
\label{HallDrude}
D_{xy}=\frac{1}{2A}\sum_{\k,n}\e_z\cdot(\j_{\k,n}^1\times\j_{\k,n}^2)\delta(\epsilon_{\k,n}-E_F)\;,
\end{align}
where $\j_{\k,n}^\ell=\langle \k,n|\j^\ell|\k,n\rangle$ and $\epsilon_{\k,n}$ and $|\k,n\rangle$ denote the eigenvalues and eigenvectors, respectively, with $\k$ inside the first Brillouin zone. $A$ labels the area of the sample and $\j^\ell$ is the current operator of layer $\ell=1,2$.

The chiral term $D_{xy}$ adds a transverse component to the longitudinal current of ordinary plasmons with opposite sign in each layer, which can be
viewed as a longitudinal magnetic moment.\cite{Stauber18} Therefore, the chiral Drude weight of Eq. (\ref{HallDrude}) endows longitudinal plasmons of twisted structures with a chiral character by linking the electric dipole oscillations $\p$ to magnetic dipole oscillations $\m$ via $\p\sim\m$. Here, we show that this property is passed onto the electromagnetic near-field and due to the strong field confinement of graphene's surface-plasmon polaritons, the field chirality is several orders of magnitude larger than the one of the corresponding far-field. Twisted atomically thin van der Waals heterostructures may thus provide the strongest near field chirality without the need of a magnetic field\cite{Poumirol17} or externally breaking the mirror plane or inversion symmetry via external nano-patterning.\cite{Fan10,Kuzyk12,Schaeferlin14}

\section*{Near-field properties of chiral plasmons in TBG} Chiral plasmonics in twisted bilayer graphene has first been introduced and discussed in the non-retarded limit,\cite{Stauber18,Stauber18b} and was recently extended including relativistic effects.\cite{Lin20} The static approach is usually enough since the transverse (s-polarised) current sources are suppressed by the fine-structure constant and can thus be neglected in comparison to the longitudinal (p-polarised) current sources. However, there are quantities which are zero in the non-retarded regime and these shall be discussed in this work within the full retarded response theory.

The plasmonic field in a general bilayer is generated by the in-plane currents $\j^\ell=j_\parallel^\ell\e_\parallel+j_\perp^\ell\e_\perp$ with $\ell=1,2$ that carry the momentum $\q=q\e_\parallel$ in the non-retarded limit. TBG displays chirality without breaking time-reversal symmetry and we have for the charged plasmon\cite{Stauber18,Stauber18b} $j_\parallel\equiv j_\parallel^1=j_\parallel^2$ and $j_\perp\equiv j_\perp^1=-j_\perp^2$. This polarization of currents is for symmetric environments and we expect it to also roughly hold in asymmetric setups such as the one for screened plasmons discussed below. 

In SI, we analyze the near-field response of one layer for which time-reversal symmetry is explicitly broken. This treatment can be extended to a bilayer with the two layers located at $z_1=a/2$ and $z_2=-a/2$ and for TBG, we set $a=3.4$\AA. An alternative approach considering an effective single electro-magnetic sheet can be found in the SI.

\subsection*{Local near-field chirality}
The formula for the local chirality and the local chirality flux couples the longitudinal and transverse field component. For real electromagnetic fields ($\Er$, $\Br$) in a dielectric medium ($\epsilon$, $\mu$), it is given by\cite{Tang10} 
\begin{align}
\label{LocalChirality}
\C&=\frac{\epsilon\epsilon_0}{2}\Er\cdot({\bm \nabla}\times\Er)+\frac{1}{2\mu\mu_0}\Br\cdot({\bm \nabla}\times\Br)\;,\\\label{ChiralityFlux}
{\bm{\mathcal F}}&=\frac{1}{2\mu\mu_0}\left(\Er\times(\nabla\times\Br)-\Br\times(\nabla\times\Er)\right)\;.
\end{align}
Both quantities are related via the continuity equation $\partial_t\C+\nabla\cdot{\bm{\mathcal F}}=0$.

We will evaluate these expressions for the near-field that is produced by the longitudinal and transverse current sources $j_\parallel$ and $j_\perp$, see the SI. 
To be more general, we will from now on explicitly consider two different dielectrics $\epsilon_i$, $\mu_i$ with $i=1,2$ in the two half-planes $|z|>a/2$ where $a$ denotes the distance between the two twisted atomic layers. 
 
In the limit $aq_i'\ll1$ with $q_i'=\sqrt{q^2-(\omega/c_i)^2}$, we then have
\begin{align}
\label{Chirality}
\C_i&=-\frac{\mu_i\mu_0}{2}aq^2j_\parallel j_\perp e^{-2q_i'|z|}\;,\\
{\bm{\mathcal F}}_i&=-\frac{\mu_i\mu_0\omega}{2}\frac{q}{q_i'}\left[2j_\parallel j_\perp q_i'a\e_\q+sqn(z)j_\parallel^2\frac{(q_i')^2}{k_i^2}\left(1+\tilde j_\perp^i\right)\e_{\q_\perp}\right]e^{-2q_i'|z|}\;,
\end{align}
where we defined $\tilde j_\perp^i=[1+\frac{j_\perp^2}{j_\parallel^2}\frac{k_i^2}{(q_i')^2}](\frac{q_i'a}{2})^2$ and $k_i=\omega/c_i$ with $c_i=c/\sqrt{\mu_i\epsilon_i}$ the speed of light of the dielectric medium. Related quantities such as the helicity and ellipticity can also be obtained from the above expressions, see SI. For completeness, let us also present the local energy density $w_i$ and the local Poynting vector ${\bf P_i}$ of each half-plane: 
\begin{align}
w_i&=\frac{\mu_i\mu_0j_\parallel^2}{2}\frac{q^2}{k_i^2}\left(1+\tilde j_\perp^i\right)e^{-2q_i'|z|}\;,\\\label{Poynting}
{\bf P}_i&=\frac{\mu_i\mu_0j_\parallel^2}{2}\frac{q\omega}{k_i^2}\left[(1+\tilde j_\perp^i)\e_\q+sqn(z)\frac{j_\perp}{j_\parallel} q_i'a\e_{\q_\perp}\right]e^{-2q_i'|z|} 
\end{align}
We note that the chirality flux as well as the Poynting vector contain a non-trivial transverse component $\e_{\q_\perp}$ which could be arbitrarily chosen without violating the respective continuity equations. The local definition of Eq. (\ref{ChiralityFlux}) and the corresponding definition of the Poynting vector thus go  beyond the transport properties as was first discussed in Ref. \cite{Berry09} .

\subsection*{Plasmon-induced chirality}
We will now discuss the plasmon-induced chirality in twisted van der Waals structures and set for simplicity $\mu_i=1$. The above equations are expressed in terms of the longitudinal and transverse current and for plasmons, they are related via $j_\perp=-2\frac{D_{xy}}{D_T}j_\parallel$.\cite{Stauber18,Stauber18b} With the knowledge of the plasmon dispersion and the linear response relation $2i\omega j_\parallel=-D_TE$ where $E$ denotes the in-plane electric field, the chirality can then be entirely written in terms of the field intensity and material constants.

If the twisted atomic layers are surrounded by two dielectrics, the intrinsic excitations are given by unscreened (optical) plasmons and their dispersion relation reads $\omega^2=qD_T/(\epsilon_0(\epsilon_1+\epsilon_2))$.\cite{Stauber14} With $E=E_{unscr}$, we then obtain for the chirality 
\begin{align}
\C_i=-\left(\frac{\epsilon_1+\epsilon_2}{2\epsilon_i}\right)^2\frac{ak_iD_{xy}}{D_T}\epsilon_i\epsilon_0k_iE_{unscr}^2e^{-2q_i'|z|}\;.
\end{align}

If the twisted atomic layers are in close proximity to a metallic plate, the plasmonic excitations are screened and their dispersion is defined by the (acoustic) sound velocity via $\omega=v_sq$. Due to the metallic gate, the chiral near-field is only present in one half-space, say $i=1$, or in between the spacer ($i=S$) and with $E=E_{scr}$, this leads to the following chirality:
\begin{align}
\C_i=-\alpha_{xy}\frac{D_T}{4v_s^2}E_{scr}^2e^{-2q_i'|z|}\;.
\end{align}
where $\alpha_{xy}=\frac{aD_{xy}}{\epsilon_0c^2}$ may be denoted as "chiral fine-structure constant". This is a small number with $\alpha_{xy}\sim5\times10^{-4}\widetilde D_{xy}$ and $D_{xy}=\widetilde D_{xy}t(e/\hbar)^2$. Nevertheless, the large local electric field $E_{scr}$ will lead to a strong chirality. 

\begin{figure}[t]
\includegraphics[width=0.99\columnwidth]{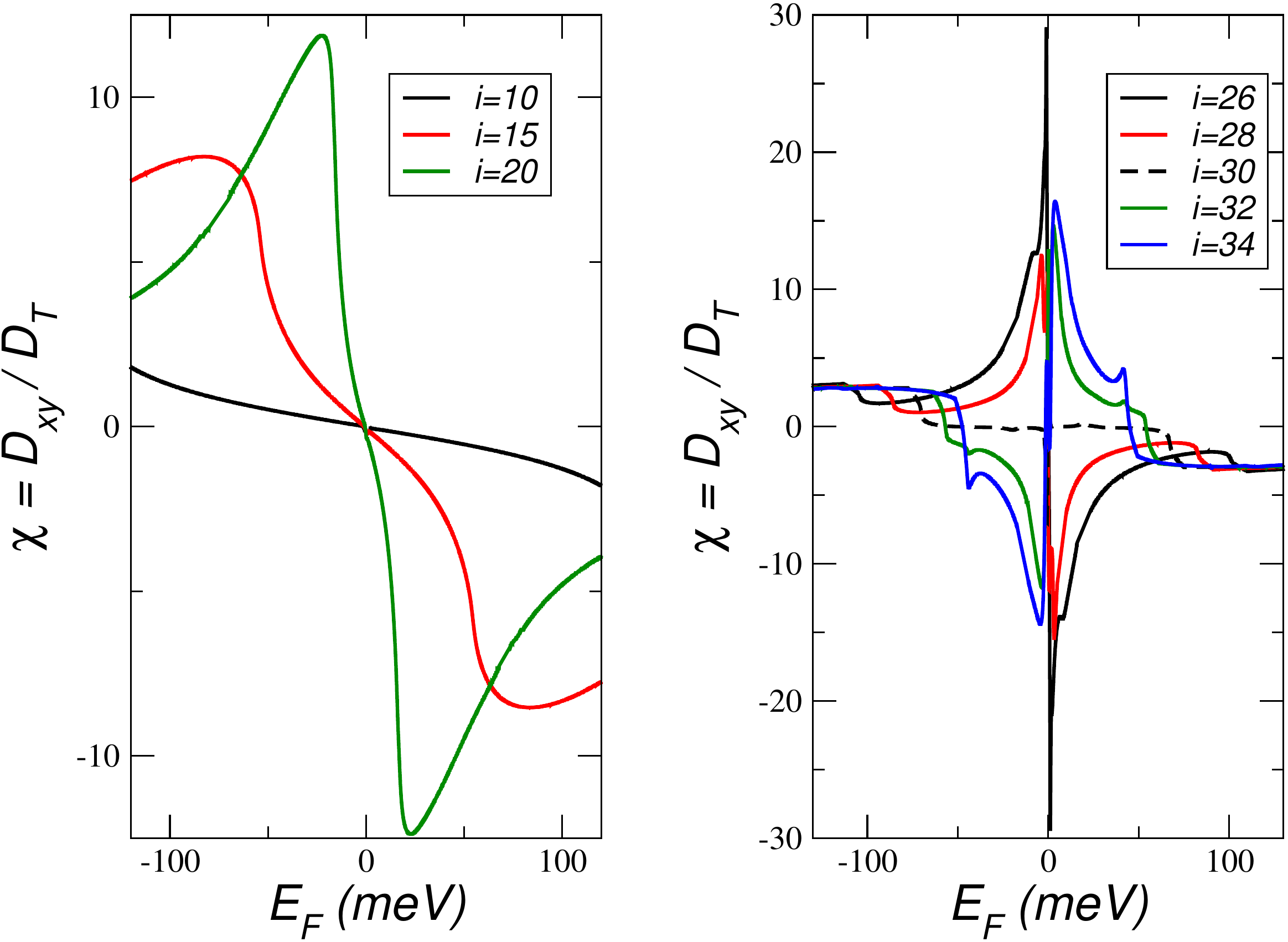}
\caption{\label{Fig3} The dimensionless quantity $\chi=D_{xy}/D_T$ for large twist angles $\theta_i=3.15^\circ,2.13^\circ,1.61^\circ$ with $i=10,15,20$ (left) and for small twist angles $\theta_i=1.25^\circ,1.16^\circ,1.08^\circ,1.02^\circ,0.96^\circ$ with $i=26,28,30,32,34$ (right) as function of the Fermi energy $E_F$. Commensurate twist angles are parametrised by $\cos\theta_i=\frac{3i^2+3i+1/2}{3i^2+3i+1}$.}
\end{figure}
\subsection*{Chiral deflection} The expressions for the local chirality depend on the total and chiral (Hall) Drude weight, shown in Figs. \ref{Fig1} and \ref{Fig2}. In Fig. \ref{Fig3}, we show the dimensionless chirality of the system $\chi=D_{xy}/D_T$ that governs the relation between the longitudinal and transverse current density and thus the local chirality of (unscreened) collective charge oscillations. It also enters in the transverse component of the Poynting vector that will lead to a chiral deflection of the energy flux. 
 
For large twist angles, $\chi$ displays an odd behavior with a maximum of around 10. For small twist angles, there is a constant plateau with well-defined chirality for large $|E_F|$ reflecting the chirality of the lattice. However, around the neutrality this chirality changes sign for $i=30$ and $D_{xy}$ becomes zero that was discussed in Ref. \cite{Stauber20} . This opens up the possibility of detecting the magic angle by pure optical means. Notice also that the curves collapse to the same value $|\chi|\sim3$ for large doping independent of the twist angle.

The analytical expressions of the previous section can be simplified considerably if only the linear term in $q'a$ is kept and retardation effects are partially neglected by $q'\to q$. The Poynting vector of a chiral plasmon then forms the angle $\tan\vartheta=-\text{sgn}(z)2\chi qa$ with respect to the propagation direction $\q$, see Eq. (\ref{Poynting}). For a twist angle of $\theta\approx2^\circ$ and a chemical potential around $\mu=40$meV, $\chi\approx10$ which would yield an angle $\vartheta\sim0.5^\circ$ that is formed by $\q$ and $ {\bm P}$. This chiral deflection should be observable via the plasmon Hall shift\cite{Shi18} at liquid-nitrogen temperatures.\cite{Ni18}

\subsection*{Comparison to the chiral far-field} Let us contrast the near-field results with the chirality obtained for left (right) CPL with $\E_i=E_0(1,\pm i,0)e^{ik_iz}$ which yields $\C_i^0=\pm \epsilon_i\epsilon_0k_i E_0^2$ and ${\bm{\mathcal F}}_i=c_i\C_i^0\e_z$. As might have been expected, the far-field chirality is thus proportional to the frequency $\omega=c_ik_i$ as well as to the field intensity. 

For unscreened (optical) plasmons, we have close to the interface ($|z|q_i\ll1$)
\begin{align}
\label{NFChiralityOp}
\left|\C_i/\C_i^0\right|={\tilde \epsilon_i}ak_i\chi F_{unscr}^2\;,
\end{align}
with the dimensionless chirality $\chi=D_{xy}/D_T$ and relative permeability ${\tilde\epsilon_i}=[(\epsilon_1+\epsilon_2)/(2\epsilon_i)]^2$. In the above formula, we have also introduced the field enhancement factor $F_{unscr}=E_{unscr}/E_0$ for unscreened plasmons.

For screened (acoustic) plasmons, we have for the upper half-space or inside the spacer ($i=1,S$)
\begin{align}
\label{NFChiralityAc}
\left|\C_i/\C_i^0\right|=\frac{\alpha_{xy}\alpha_T}{4\epsilon_idk_i}F_{scr}^2\;,
\end{align}
where $\alpha_{T}=\frac{dD_{T}}{\epsilon_0v_s^2}$ is a constant. We have further introduced the spacer distance $d$ for convenience and the field enhancement factor $F_{scr}=E_{scr}/E_0$ for screened plasmons. 

In the following, we will differentiate between moderately and strongly screened plasmons. In the case of moderately screened plasmons, the sound velocity is larger than the Fermi velocity $v_s\gtrsim v_F$. It depends on the total Drude weight and is given by $v_s^2=dD_T/(\epsilon_0\epsilon_S)$ which yields $\alpha_T=\epsilon_S$.\cite{Stauber12} Note that in this case the chirality depends inversely on the spacer distance $d$. For strongly screened plasmons, the sound velocity approaches the Fermi velocity $v_s\to v_F$ which sets the lower bound imposed by the non-local charge response of Dirac electrons.\cite{Wunsch06,Stauber2012} In this case, the chirality does not depend on the distance $d$ anymore as it has reached the physical confinement limit. 

Eqs. (\ref{NFChiralityOp}) and (\ref{NFChiralityAc}) would be an artifact unless we relate both $E=E_{unscr,scr}$ and $E_0$ by a common physical ruler. For that, we imagine that the same plasmonic current intensity $j_{\parallel}$ that creates $E$, now shines at a radiative wavevector $q_r\ll\omega/c$, creating the far-field plane wave of amplitude $E_0$. Then $F_{scr}\sim \tfrac{q}{\omega/c}\sim \tfrac{c}{v_s}$, valid for the typical wavevectors of graphene plasmons which will be used in the following. $F_{scr}$ is usually larger than the one for unscreened plasmons. In an experimental setup, this can be controlled with a proximal metal plate to the TBG separated by a thin dielectric spacer. When the spacer thickness approaches $1\,$nm, the field confinement for the screened acoustic plasmon can be $\sim 5$ times larger than that of the unscreened case \cite{lee2019graphene}.
In the following, we will thus set $F_{unscr}=0.2\tfrac{c}{v_F}$.

\begin{figure}[t]
\includegraphics[width=0.99\columnwidth]{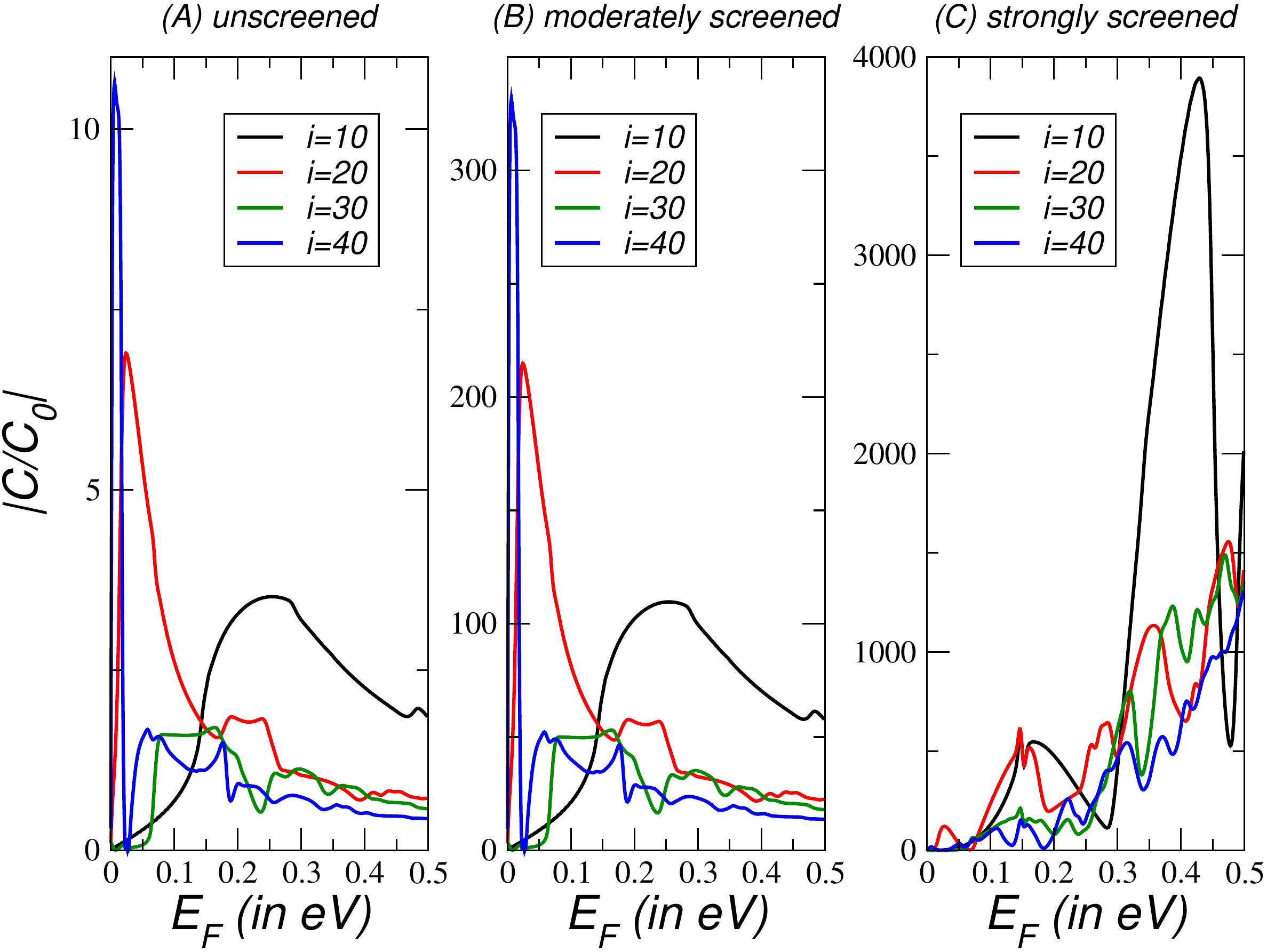}
\caption{\label{Fig4} Near-field chirality for different plasmonic regimes normalised by the corresponding far-field chirality of circularly polarized light and different twist angles $\theta_i=3.15^\circ,1.61^\circ,1.08^\circ,0.82^\circ$ with $i=10,20,30,40$ as function of the Fermi energy $E_F$. Commensurate twist angles are parametrised by $\cos\theta_i=\frac{3i^2+3i+1/2}{3i^2+3i+1}$.}
\end{figure}

\subsection*{Estimating near-field chirality}
In Fig. \ref{Fig4}, we plot the normalised near-field chirality with respect to the corresponding far-field chirality of circularly polarized light $\left|\C/\C^0\right|$ for different twist angles as function of the Fermi energy $E_F$ for the half-space $i=1$. This half-space is assumed to be free space with $\epsilon_1=1$. We also set $\epsilon_2=\epsilon_S=5$ modelling a hBN-substrate used in typical experiments.\cite{Cao18b,Yankowitz19,Lu19} In the case of screened plasmons, we further set the spacer width to be $\sim5$ hBN-layers, i.e, $d=5a\sim17$\AA. Finally, we set $\lambda_1=2\pi/k_1=10\mu$m used in typical scanning near-field experiments,\cite{Sunku18} but in the SI, also other parameters are discussed.

In Fig. \ref{Fig4}(A), we plot the enhancement factor for unscreened plasmons which turns out to be of the order of 10. It can further be enlarged by increasing the plasmonic frequency. For moderately screened plasmons, the sound velocity $v_s\gtrsim v_F$ is given by $v_s^2=dD_T/(\epsilon_0\epsilon_S)$.\cite{Stauber12} Since the field enhancement now becomes $(c/v_s)^2\sim D_T^{-1}$, we obtain the same functional dependence as in the case of unscreened plasmons. However, the enhancement factor turns out to be larger by a factor $\sim30$ for $d=5a$, as shown in Fig. \ref{Fig4}(B). 

For strongly screened plasmons, the sound velocity approaches the Fermi velocity and we set $v_s=v_F$.\cite{Alcaraz18} This yields the largest enhancement factor of the order of $10^3$ for twist angles $\theta\sim3^\circ$ and for large electronic density with $E_F\sim0.3-0.5$eV as shown in Fig. \ref{Fig4}(C). Highly doped twisted van der Waals heterostructures screened by a nearby metallic gate should thus yield the strongest chiral near-field without breaking time-reversal symmetry. Moreover, the chirality is relatively broadband and highly tunable through carrier density modulation via gate voltage. Further enhancement might be possible due to the Fermi velocity renomalization by replacing $v_F\to v_F^*$ with $v_F^*<v_F$.\cite{Lopes07,Bistritzer11} Finally, let us also mention the possibility to place TBG in a cavity that should enhance the chirality at resonant frequencies.

\subsection*{Chiral Chemistry}
Strong light-matter interaction can induce or catalize new reactions. E.g., there are interesting proposals to enhance two-photon processes\cite{Rivera17} and to construct "designer atoms".\cite{Chang17} 
They rely on the wavelength reduction of confined, plasmonic excitations. An alternative, but related approach to alter chemical reactions or to catalyse new ones is to drive the system into the strong light-matter interaction regime.\cite{Hutchison12,Flick17,Galego17,Flick18,Feist18,Yuen-Zhou19} 
We shall add to this general approach an additional degree of freedom, namely, the chirality of the near-field.

For screened plasmons, which can be achieved with a proximal metal gate,\cite{lee2019graphene,alonso2017acoustic}
the chiral enhancement is largest for twist angles of $\theta\sim3^\circ$ and for large electronic density $n\sim10^{13}$cm$^{-1}$. This would yield an asymmetry factor in the circular dichroism of order unity and the proposed platform might give rise to unprecedented chemical reactions between chiral molecules that are usually forbidden in the spirit of previous proposals, see Fig. \ref{Fig4}.\cite{Rivera17,Chang17,Hutchison12,Flick17,Galego17,Flick18,Feist18,Yuen-Zhou19} 

Let us finally note that $\C$ also denotes the chiral selectivity of the near-field coupling with an emitter, and thus can also modify the polarized photoluminescence of dye molecules.\cite{meinzer2013probing} The adsorbed molecules which can effectively couple far-field light into the plasmons, would have different near-field coupling efficiencies for the two enantiomers.\cite{Hentschel17} Since the electronic chirality changes sign at the magic angle,\cite{Stauber20} this selectivity can also be used to detect the magic angle by pure optical means.

\section*{Summary and discussion}  
We have investigated the electromagnetic near-field confined to TBG focusing on its chirality. The rotational mismatch breaks all mirror plane symmetries and chirality arises from the quantum nature of interlayer coupling. The effect does thus not rely on the breaking of  time-reversal symmetry and can, therefore, be useful in the context of catalysing chemical reactions without changing the external conditions, e.g., due to the presence of a magnetic field. We find huge field enhancements especially in the case of acoustic plasmons paving the way towards {\it chiral  plasmon-induced chemistry} which is possible for general {\it twisted} van der Waals structures\cite{Geim13} since the interlayer Moir\'e coupling induces a chiral response that endows surface plasmons with a chiral character.\cite{Stauber18,Stauber18b,Lin20} Unlike most proposals in the field of chiral plasmonics \cite{Hentschel17}, the near-field chirality discussed herein has a quantum origin due to the interlayer coupling between the atomic layers, and does not rely on any nanofabrication of metallic chiral structures.

Our proposal can be extended to other 2D materials beyond graphene, and provides a novel approach to macroscopic chiral platforms for optics, sensing, chemistry and beyond. One example might concern a variety of synthetic methods for amino acids that lead to equal amounts of left- and right-handed enantiomers, i.e., they are usually obtained as a racemate and processes for enantiomer separation must then be carried out if pure L- or D-amino acids are required. Performing the synthesis close to van der Waals heterostructure and exciting plasmons with a well-defined chirality might lead to enantiomerically pure L or D amino acids. Also, entirely new catalysed reaction of chiral molecules can be envisioned.

\section*{Acknowlegdements}
We thank J. Gonz\'alez and J. Schliemann for discussions. This work has been supported by Spain's MINECO under Grant No. FIS2017-82260-P, PGC2018-096955-B-C42, and CEX2018-000805-M as well as by the CSIC Research Platform on Quantum Technologies PTI-001. TS also acknowledges support from the "Salvador de Madariaga"-Programme under Grant No. PRX19/00024 and from Germany's Deutsche Forschungsgemeinschaft (DFG) via SFB 1277. TL acknowledges support by the National Science Foundation, NSF/EFRI Grant No. EFRI-1741660.

\newpage
{\bf\huge Supplemental Information}

\section*{Continuum model of TBG and linear response}
In the following, we will perform the detailed calculations for the continuum model as proposed in Refs. \cite{Lopes07,Bistritzer11} . For one valley, the Hamiltonian reads
\begin{align}
H=
\hbar v_F \left(
 \begin{array}{cccc}
0 &  -i\partial_x - \partial_y + i \frac{\Delta{\bf K}}{2} & 
                           V_{AA'}(\mathbf{r}) & V_{AB'}(\mathbf{r}) \\
 -i\partial_x + \partial_y - i \frac{\Delta{\bf K}}{2} & 0 & 
                           V_{BA'}(\mathbf{r}) & V_{AA'}(\mathbf{r}) \\
 V_{AA'}^\star(\mathbf{r}) & V_{BA'}^\star(\mathbf{r}) & 0 & 
                         -i\partial_x - \partial_y - i \frac{\Delta{\bf K}}{2} \\
 V_{AB'}^\star(\mathbf{r}) & V_{AA'}^\star(\mathbf{r}) &  
                        -i\partial_x + \partial_y + i \frac{\Delta{\bf K}}{2} & 0
 \end{array}\right)\;,
\label{H}
\end{align}
where we introduced the Fermi velocity of graphene $v_F$, the shift between the two Dirac cones $\Delta{\bf K}$, and $V_{AA'}({\bf r}), V_{AB'}({\bf r})$, and $V_{BA'}({\bf r})$ are the respective interlayer tunneling amplitudes between regions of stacking $AA, AB$, and $BA$. The Hamiltonian for the opposite valley is obtained by replacing $\Delta \mathbf{K}$ by $-\Delta \mathbf{K}$ and reversing the momentum $k_x$.

Twist angles shall be commensurate and are parametrised by $\cos\theta_i=\frac{3i^2+3i+1/2}{3i^2+3i+1}$. We also set the in-plane hopping parameter $t=2.78$eV which is related to the Fermi velocity via $v_F=\frac{\sqrt{3}}{2}ta_0$ with $a_0=2.46$\AA. The interlayer hopping strength is taken as $w=0.11$meV following the  notation of Ref. \cite{Bistritzer11}.

The conductivity $\bm \sigma$ shall also be defined by a $4\times4$ matrix  with
\begin{equation}\label{deltadiag}
\begin{bmatrix} \bm j^{(1)} \\  \bm j^{(2)}  \end{bmatrix} = \bm \sigma
\begin{bmatrix} \bm E^{(1)} \\  \bm E^{(2)}  \end{bmatrix} 
,\end{equation}
where $\bm j^{\ell} $ and $\bm E^{\ell}$  represent in-plane currents and total fields in the plane $\ell=1,2$. For a rotationally invariant, symmetric system, we can then write the response in the following way \cite{Stauber18}:
\begin{align}
\bm \sigma=
\left(
 \begin{array}{cccc}
\sigma_0  &  0 & 
                            \sigma_1 & \sigma_{xy} \\
0 & \sigma_0 & 
                           -\sigma_{xy} & \sigma_1 \\
\sigma_1 & -\sigma_{xy} & \sigma_0  & 
                         0 \\
\sigma_{xy} & \sigma_1&  
                        0& \sigma_0
 \end{array}\right)\;,
\label{sigma}
\end{align}
where $\sigma_{0,1}(\omega) $, $\sigma_{xy}(\omega)$  are complex functions  characterizing the local in-plane response. They can be interpreted as the in-plane conductivity, the covalent drag conductivity as well as the Hall or chiral conductivity, respectively. 

It is usually sufficient to discuss plasmonic excitations with respect to their Drude weights and the Drude weight $D_\nu$ is related to the low-frequency limit of the response function $\sigma_\nu$ via $D_\nu=\lim_{\omega\to0}\omega$Im$\sigma_\nu$ with $\nu=0,1,xy$. The total Drude weight is then obtained by $D_T=2(D_0+D_1)$, and it is also given by the familiar inverse mass formula
\begin{align}
D_{T}=\frac{1}{A}\sum_{\k,n}\left(\frac{e}{\hbar}\frac{\partial\epsilon_{\k,n}}{\partial k_x}\right)^2\delta(\epsilon_{\k,n}-E_F),
\end{align}
which only needs the knowledge of the band-structure of the Hamiltonian Eq. (\ref{H}). For the chiral response, we will use Eq. (1) of the main text \cite{Stauber20}. Both quantities characterise the chiral near-field and are plotted and discussed in the main text. 

\section*{Near-field properties in 2D systems with broken time-reversal symmetry}
\label{BrokenTR}
Two-dimensional bulk plasmonic properties in systems with time-reversal symmetry cannot easily be distinguished from systems with broken time-reversal symmetry regarding their dispersion. The reason for that is that the transverse (s-polarised) current sources are suppressed by the fine-structure constant and can thus be neglected in comparison to the longitudinal (p-polarised) current sources. Still, there are quantities which are only non-zero in the retarded regime such as the helicity, ellipticity or chirality.

Plasmons in two-dimensional systems with broken time-reversal symmetry exhibit a transverse current $j_\perp$ that is associated to the longitudinal current $j_\parallel$. Let us define the complex sources $\j_\parallel=j_\parallel\e_\q e^{i\q\cdot\br}e^{-2q'|z|}$ and $\j_\perp=j_\perp\e_{\q_\perp} e^{i\q\cdot\br}e^{-2q'|z|}$ with $q'=\sqrt{q^2-\mu\epsilon(\omega/c)^2}$ and $\e_{\q_\perp}=\e_z\times\e_\q$. The corresponding real current densities shall be given by ${\bf \mathfrak{j}}_\nu=$Re$\j_\nu$ with $\nu=\parallel,\perp$. 

The associated near-field is given by $\A_\nu=-\mathcal{D_\nu}\j_\nu$ where $\mathcal{D_\nu}$ is the longitudinal ($\nu=\parallel$) or transverse ($\nu=\perp$) photonic propagator, respectively.\cite{Stauber12} The parallel current $j_\parallel$ will thus give rise to a longitudinal field and the perpendicular current $j_\perp$ to a transverse field. With the total gauge field $\A=\A_\parallel+\A_\perp$ and $\E=i\omega\A$ and $\B=\nabla\times\A$, we thus obtain the following expression:
\begin{align}
\E&=i\omega\left(
\begin{array}{c}
-d_lj_\parallel\\
-d_tj_\perp\\
-isgn(z)\frac{q}{q'}d_lj_\parallel
\end{array}\right)e^{iqx}e^{-q'|z|}\\
\B&=\left(
\begin{array}{c}
-sgn(z)q'd_tj_\perp\\
-sgn(z)\frac{k_0^2}{q'}d_lj_\parallel\\
-iqd_tj_\perp
\end{array}\right)e^{iqx}e^{-q'|z|}\;,
\end{align}
where $d_l=\frac{q'}{2\epsilon\epsilon_0\omega^2}$, $d_t=-\frac{\mu\mu_0}{2q'}$, $k_0=\omega/c$ and $\e_\q=\e_x$.

\subsection*{Optical momentum, spin, angular momentum, and helicity}
The {\it local} energy density $w$, "complex" Poynting vector $\Pi$ (the Poynting vector is defined as $\P_{Poy}=\Re\Pi$), momentum $\P$, spin $\S$, and helicity $\mathcal{H}$ of a monochromatic electromagnetic wave can be defined as follows:\cite{Berry09,Bliokh14,Bliokh14b,Bliokh15,Bliokh17}
\begin{align}
w&=\frac{\epsilon\epsilon_0}{4}\E^*\cdot\E+\frac{1}{4\mu\mu_0}\B^*\cdot\B\\
{\bm \Pi}&=\frac{1}{2\mu\mu_0}\E^*\times\B\\
\omega\P&=\frac{\epsilon\epsilon_0}{4}\Im\E^*\cdot(\nabla)\E+\frac{1}{4\mu\mu_0}\Im\B^*\cdot(\nabla)\B\\
\omega\S&=\frac{\epsilon\epsilon_0}{4}\Im\E^*\times\E+\frac{1}{4\mu\mu_0}\Im\B^*\times\B\\
\omega \mathcal{H}&=-\frac{1}{2\mu\mu_0}\Im\E^*\cdot\B
\end{align}

For the above current density, this yields the following local properties with $\q=q\e_x$:
\begin{align}
w&=\frac{\mu\mu_0j_\parallel^2}{8}\frac{q^2}{k_0^2}\left(1+\frac{j_\perp^2}{j_\parallel^2}\frac{k_0^2}{(q')^2}\right)e^{-2q'|z|}\\
{\bm \Pi}&=\frac{\mu\mu_0j_\parallel^2}{8}\frac{q\omega}{k_0^2}\left(
\begin{array}{c}
1+\frac{j_\perp^2}{j_\parallel^2}\frac{k_0^2}{(q')^2}
\\
sgn(z)2\frac{j_\perp}{j_\parallel} \\
-sgn(z)i\frac{q'}{q}\left(1-\frac{j_\perp^2}{j_\parallel^2}\frac{k_0^2}{(q')^2}\right)
\end{array}\right)e^{-2q'|z|}\\
\P&=\frac{q}{\omega}w\e_\q e^{-2q'|z|}\\
\S&=-sgn(z)\frac{\mu\mu_0j_\parallel^2}{8}\frac{1}{\omega}\frac{q}{q'}\left(
\begin{array}{c}
2\frac{j_\perp}{j_\parallel}\\
\frac{(q')^2}{k_0^2}\left(1+\frac{j_\perp^2}{j_\parallel^2}\frac{k_0^2}{(q')^2}\right)
\\
0
\end{array}\right)e^{-2q'|z|}\\
\mathcal{H}&=-sgn(z) \frac{\mu\mu_0j_\parallel j_\perp}{4}\frac{q^2}{q'k_0^2}e^{-2q'|z|}
\end{align}
\subsection*{Ellipticity}
We can also define the ellipticity of the electric and magnetic field as follows
\begin{align}
\mathcal{E}_\E&=\Im\frac{\E^*\times\E}{\E^*\cdot\E}\cdot\P_{Poy}\;,\\
\mathcal{E}_\B&=\Im\frac{\B^*\times\B}{\B^*\cdot\B}\cdot\P_{Poy}\;,\\
\mathcal{E}&=\Im\frac{\epsilon\epsilon_0\E^*\times\E+\frac{1}{\mu\mu_0}\B^*\times\B}{\epsilon\epsilon_0\E^*\cdot\E+\frac{1}{\mu\mu_0}\B^*\cdot\B}\cdot\P_{Poy}\;,
\end{align}
with $\P_{Poy}=\Re\Pi$. Interestingly, we find $\mathcal{E}_\E=\mathcal{E}_\B$ with
\begin{align}
\mathcal{E}_\E=-sgn(z)\frac{\mu\mu_0j_\parallel j_\perp}{4}\frac{\omega}{k_0^2}\frac{q^2}{q'}\;.
\end{align}
Even though the electric and magnetic fields are coupled, we can assign the same ellipticity to each sector. Consequently, we also have $\mathcal{E}=\mathcal{E}_\E$. Furthermore, there is a relation between the ellipticity and helicity with $\mathcal{E}=\omega\mathcal{H}$.
\subsection*{Chirality}
The local chirality for a real electromagnetic field in a dielectric medium ($\epsilon$, $\mu$) is defined by\cite{Tang10}
\begin{align}
\C=\frac{\epsilon\epsilon_0}{2}\Er\cdot({\bm \nabla}\times\Er)+\frac{1}{2\mu\mu_0}\Br\cdot({\bm \nabla}\times\Br)\;.
\end{align}
Note that a finite contribution to the chirality only comes from the scalar product involving both, the longitudinal and the transverse field component which justifies the denomination of this conserved quantity. The resulting fields, therefore, display the following chirality:
\begin{align}
\C=-sgn(z)\frac{\mu\mu_0}{4}\frac{j_\parallel j_\perp q^2}{q'} e^{-2q'|z|}\;
\end{align}

The optical chirality is associated to a flux of chirality related by the usual continuity equation $\partial_t\C+\nabla\cdot{\bm{\mathcal F}}=0$ (in the absence of material currents) which is locally defined as
\begin{align}
{\bm{\mathcal F}}=\frac{1}{2\mu\mu_0}\left(\Er\times(\nabla\times\Br)-\Br\times(\nabla\times\Er)\right)\;.
\end{align}
For the chiral plasmon, this gives
\begin{align}
{\bm{\mathcal F}}&=-sgn(z)\frac{\mu\mu_0\omega}{8}\frac{q}{q'}e^{-2q'|z|}\left[2j_\parallel j_\perp \e_\q+j_\parallel^2\frac{(q')^2}{k_0^2}\left(1+\frac{j_\perp^2}{j_\parallel^2}\frac{k_0^2}{(q')^2}\right)\e_{\q_\perp}\right]\;.
\end{align}

Notice that we have ${\bm{\mathcal F}}=\omega^2\S$ which gives rise to a conserved "spin-density" $\mathcal{S}=\C/\omega^2$. We also have $\mathcal{H}=\mathcal{C}/k_0^2$. Both relations are general and obtained by noting that 
\begin{align}
\mathcal{C} &= \frac{\epsilon\epsilon_0}{2} (\Br \cdot \partial_t \Er - \Er \cdot \partial_t \Br)=k_0^2\mathcal{H}\;,\\
{\bm{\mathcal F}} &= \frac{\epsilon\epsilon_0}{2} \Er \times \partial_t \Er +\frac{1}{2\mu\mu_0}\Br \times \partial_t \Br=\omega^2{\bm S}\;.
\end{align}
The chirality is also linked to the ellipticity with $\mathcal{E}=\omega\C/k_0^2$. 

\section*{Near-field properties in 2D systems with time-reversal symmetry}
In the previous Section, we have analyzed the near-field response in one layer for which time-reversal symmetry is explicitly broken. This treatment can be extended to a bilayer system with the two layers located at $z_1=a/2$ and $z_2=-a/2$ and with opposing current densities. The whole system thus does not break time-reversal  symmetry as indicated by the symmetric response matrix Eq. (\ref{sigma}). An alternative approach considering an effective single electro-magnetic sheet can also be found in the last Section.

Let us now also explicitly consider two different dielectrics $\epsilon_i$, $\mu_i$ with $i=1,2$ in the two half-planes $|z|>a/2$. The electromagnetic field is usually characterized by the local energy density $w_i$ and the local Poynting vector ${\bf P_i}$ of each half-plane. In the limit $aq_i'\ll1$ with $q_i'=\sqrt{q^2-\mu_i\epsilon_i(\omega/c)^2}$, we obtain the following expressions:
\begin{align}
w_i&=\frac{\mu_i\mu_0j_\parallel^2}{2}\frac{q^2}{k_i^2}\left(1+\tilde j_\perp^i\right)e^{-2q_i'|z|}\;,\\
{\bf P}_i&=\frac{\mu_i\mu_0j_\parallel^2}{2}\frac{q\omega}{k_i^2}\left[(1+\tilde j_\perp^i)\e_\q+sqn(z)\frac{j_\perp}{j_\parallel} q_i'a\e_{\q_\perp}\right]e^{-2q_i'|z|}\;,
\end{align}
where we defined $\tilde j_\perp^i=[1+\frac{j_\perp^2}{j_\parallel^2}\frac{k_i^2}{(q_i')^2}](\frac{q_i'a}{2})^2$ and $k_i=\omega/c_i$ with $c_i=c/\sqrt{\mu_i\epsilon_i}$ the speed of light of the dielectric medium.

For the near-field chirality  of TBG, we have as stated in the main text
\begin{align}
\label{Chirality}
\C_i&=-\frac{\mu_i\mu_0}{2}aq^2j_\parallel j_\perp e^{-2q_i'|z|}\;,\\
{\bm{\mathcal F}}_i&=-\frac{\mu_i\mu_0\omega}{2}\frac{q}{q_i'}\left[2j_\parallel j_\perp q_i'a\e_\q+sqn(z)j_\parallel^2\frac{(q_i')^2}{k_i^2}\left(1+\tilde j_\perp^i\right)\e_{\q_\perp}\right]e^{-2q_i'|z|}\;.
\end{align}
Notice that the Poynting vector as well as the chirality flux contain a non-trivial transverse component which could be chosen arbitrarily without violating the continuity equation. The local definition of $\P$ and ${\bm{\mathcal F}}$ thus goes beyond the transport properties.\cite{Berry09}

\section*{Comparison to far-field chirality: Figure of merit}
We will now discuss the plasmon-induced chirality in twisted van der Waals structures and as in the main text, we set $\mu_i=1$ for simplicity. The above equations are expressed in terms of the longitudinal and transverse current and for plasmons, they are related via $j_\perp=-2\frac{D_{xy}}{D_T}j_\parallel$ \cite{Stauber18,Stauber18b}. With the knowledge of the plasmon dispersion and the linear response relation $2i\omega j_\parallel=-D_TE$ where $E$ denotes the in-plane electric field, the chirality can then be entirely written in terms of the field intensity:
\begin{align}
\label{ChiralityField}
\C_i&=-\frac{a\mu_0}{4}\frac{q^2}{\omega^2}D_{xy}D_TE^2e^{-2q_i'|z|}
\end{align}

This expression depends on the ratio of $q/\omega$ and thus on the nature of the plasmonic excitation. Assuming an upper and lower dielectric with $\epsilon_1$ and $\epsilon_2$, respectively, leads to the 2D dispersion relation for optical (unscreened) plasmons, $\omega^2=qD_T/(\epsilon_0(\epsilon_1+\epsilon_2))$. With $c_i^{-2}=\epsilon_i\epsilon_0\mu_0$ and $\omega=c_ik_i$, this then yields
\begin{align}
\C_i=-\left(\frac{\epsilon_1+\epsilon_2}{2\epsilon_i}\right)^2\frac{ak_iD_{xy}}{D_T}\epsilon_i\epsilon_0k_iE^2e^{-2q_i'|z|}\;.
\end{align}

Assuming a metallic gate at the distance $d$ from the TBG, the plasmonic excitations become screened and obey a linear dispersion relation $\omega=v_sq$, characterized by the sound velocity $v_s$. The electric field is reduced by a factor $(1-e^{-2q_S'd})$ due to destructive interference, but at the same time strongly enhanced due to the field confinement. We will account for both effects by an effective field intensity. This yields
\begin{align}
\C_i&=-\frac{a}{4\epsilon_0}\frac{D_{xy}D_T}{c^2v_s^2}E^2e^{-2q_i'|z|}\;.
\end{align}

Let us now introduce a dimensionless figure of merit by comparing the near-field chirality with the corresponding far-field chirality given by $\C_i^0=\pm \epsilon_i\epsilon_0k_i E_0^2$. For unscreened plasmons, we then obtain  
\begin{align}
\label{NFChiralityOp}
\left|\C_i/\C_i^0\right|=\left(\frac{\epsilon_1+\epsilon_2}{2\epsilon_i}\right)^2ak_i\frac{D_{xy}}{D_T}\left(\frac{E}{E_0}\right)^2\;,
\end{align}
that reproduces Eq. (10) of the main text.

For screened (acoustic) plasmons, we have for the upper half-space or inside the spacer ($i=1,S$)
\begin{align}
\label{NFChiralityAc}
\left|\C_i/\C_i^0\right|=\frac{aD_{xy}}{\epsilon_0c^2}\frac{dD_{T}}{\epsilon_0v_s^2}\frac{1}{4\epsilon_idk_i}\left(\frac{E}{E_0}\right)^2\;.
\end{align}

In Fig. \ref{Fig4},we show the near-field chirality for different dielectric constants $\epsilon_1=1,5$ and different (free space) wavelengths $\lambda=1,10\mu$m as function of the Fermi energy $E_F$. Whereas a large dielectric suppresses the overall chirality, changing the wavelength has opposite effects with respect to (A) unscreened and (B) screened plasmons.

\begin{figure}[t]
\includegraphics[width=0.89\columnwidth]{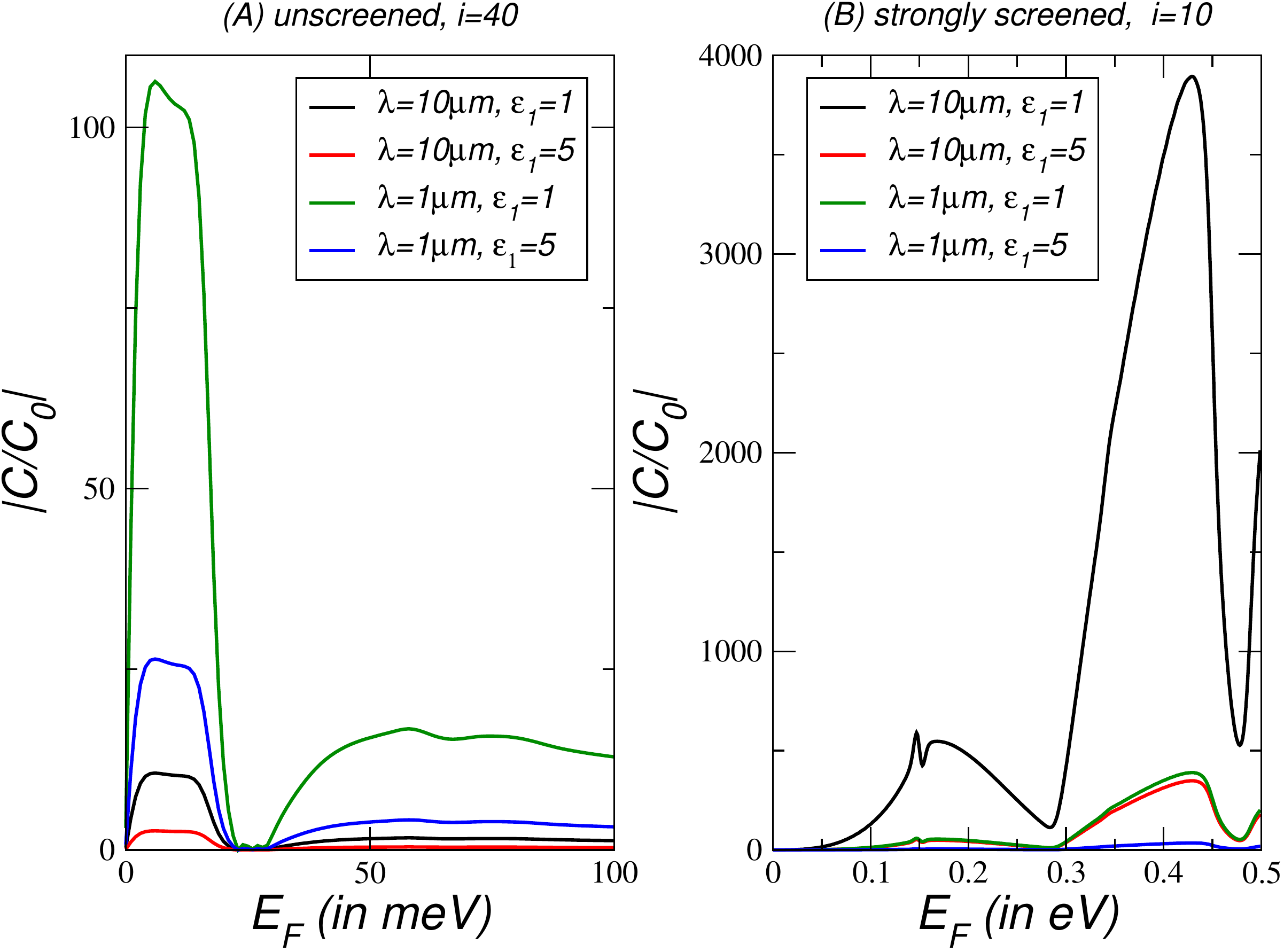}
\caption{\label{Fig4} Near-field chirality for different plasmonic regimes normalised by the corresponding far-field chirality of circularly polarized light for different dielectric constants $\epsilon_1=1,5$ and different (free space) wavelengths $\lambda=1,10\mu$m as function of the Fermi energy $E_F$. (A) unscreened plasmons for  $\theta_i=0.82^\circ$ with $i=40$.  (B) strongly screened plasmons for $\theta_i=3.15^\circ$ with $i=10$.}
\end{figure}

\section*{Chirality in magneto-electric sheets}
\label{MagneticSheet}
In this Appendix, we will consider an effective magneto-electric sheet. For chiral plasmons, the classical picture of field sources in vacuum are then in-plane, longitudinal current and magnetic moment densities,  written as 
\begin{equation}
\begin{split}
2\bm j(\bm r,t ) &= \bm j_0 \, e^{i \bm q \cdot \bm r} e^{-i \omega t}  + c.c. \\
2\bm m(\bm r,t ) &=  a \tilde{ m}\, \bm j_0\, e^{i \bm q \cdot \bm r} e^{-i \omega t}  + c.c.
\end{split}
,\end{equation}
with $\bm j_0 \parallel \bm q $,  and  where $\tilde{m}$ is a  material constant that quantifies the parallel magnetic moment  following the current. $a$ represents an intrinsic length (interlayer distance for twisted bilayer), introduced to make $\tilde{m}$ dimensionless.  

The fields associated with these sources, $\bm E_{\bm j,\bm m}$ and $\bm B_{\bm j,\bm m} $,   can be calculated explicitly  but, to show that $\bar{\mathcal{C}}$ is non zero, and the generality of the argument, it suffices to realize that: i) only the crossed terms $(\bm j,\bm m) $ contribute to $\bar{\mathcal{C}}$ on symmetry (parity) grounds, ii) the term from $\bm E_{\bm m}  \bm B_{\bm j}$ is smaller than the term $\bm E_{\bm j}  \bm B_{\bm m}$ by factors of  $(\tfrac{\omega}{q c})^2  $, iii) the electric and magnetic fields of an electric dipole can be read from the magnetic and electric counterparts of a magnetic dipole.

The last point and the relation between current and dipole density, $\bm j = \partial_t \bm p $, allow us to write
\begin{equation}
\bm B_{\bm m} = i \omega \epsilon\epsilon_0 \mu\mu_0 \tilde{m} a \bm E_{\bm j}
,\end{equation}
and finally
\begin{equation}\label{final}
\mathcal{C} =  k_0^2  \frac{\epsilon\epsilon_0}{4} (\tilde{m}+\tilde{m}^*) a |\bm E_{\bm j} |^2
.\end{equation}
We recall that  $\bm E_{\bm j}$ is the electric field associated with the longitudinal plasmon current and, for the near field, one could safely take the instantaneous approximation for it.  Its explicit expression in terms of $\bm j_0$ will lead to the standard exponential decay $e^{-2 q |z|}$ of near fields.

As a final remark, one notices that a real value of $\tilde{m}$ is required for finite  $\bar{\mathcal{C}}$. This implies that the magnetic moment has to have a component in phase with the plasmon longitudinal current and, therefore, the magnetic dipole density is in {\em quadrature} with the electric dipole density. This is precisely the condition for an atomic transition to be chirally active, so the whole picture is consistent. Furthermore, although the formalism is tailored to layered systems, the generality of the arguments implies that  a chiral near field should exist whenever the sources comply with the previous requirement of parallel and in quadrature electric and magnetic moments.
\bibliography{Chirality2.bib} 
\end{document}